# Evidence for a Two-stage Melting Transition of the Vortex Matter in $Bi_2Sr_2Ca_1Cu_2O_{8+\delta}$ Single Crystals obtained by Muon Spin Rotation

T. Blasius[1], Ch. Niedermayer[1], J.L. Tallon[2], D.M. Pooke[2], A. Golnik[3], and C. Bernhard[3]

[1]Universität Konstanz, Fakultät für Physik, D-78434 Konstanz, Germany
[2]New Zealand Institute for Industrial Research, P.O. Box 31310 Lower Hutt, New Zealand
[3]Max-Planck-Institut für Festkörperforschung, D-70569 Stuttgart, Germany

From muon spin rotation measurements on under- to overdoped Bi-2212 crystals we obtain evidence for a two-stage transition of the vortex matter as a function of temperature. The first transition is well known and related to the irreversibility line (IL). The second one is located below the IL and has not been previously observed. It occurs for all three sets of crystals and is unrelated to the vortex mobility. Our data are consistent with a two-stage melting scenario where the intra-planar melting of the vortex lattice and the inter-planar decoupling of the vortex lines occur independently.

PACS numbers: 74.60.Ge, 74.25.Dw, 74.72.Hs, 76.75.+i

The vortex matter in cuprate high-$T_c$ superconductors exhibits a complex phase diagram as a function of magnetic field, temperature and doping [1 - 3]. With the magnetic field applied perpendicular to the $CuO_2$ planes (H//c) the flux-lines can be viewed as stacks of pancake vortices that reside within the superconducting planes and are weakly coupled across the insulating spacer layers mainly via Josephson tunneling [4]. Both the intra-planar vortex order and the inter-planar coupling between the pancake vortices forming a flux-line can be overcome by thermal fluctuations. In principle, these processes may occur independently and at different temperatures, i.e. the intra-planar melting of the flux-line lattice (FLL) may take place at a lower temperature than the inter-planar decoupling of the individual flux-lines [5]. Experimental evidence for such a two-stage melting scenario has indeed been obtained from transport measurements [6, 7] on the highly anisotropic system $Bi_2Sr_2CaCu_2O_{8+\delta}$ (Bi-2212). These experiments, however, only probe the vortex-state under non-equilibrium conditions.

The technique of transverse-field muon spin rotation (TF-µSR) is an ideal tool to study the equilibrium vortex state. In this letter we present the results of such a study on three sets of overdoped ($T_c$ = 64 K), nearly optimized ($T_c$ = 90 K) and underdoped ($T_c$ = 77 K) Bi-2212 single crystals which provide the first experimental indication for a two-stage transition of the vortex matter under equilibrium conditions. We discuss the evidence that the observed changes in the vortex-state correspond to a two-stage melting transition of the vortex matter.

The under- and overdoped sets of crystals were grown by a floating-zone technique as described elsewhere [8]. The oxygen content was achieved by annealing in a gas stream of 0.02 % $O_2$ at 765 °C followed by rapid quenching for the underdoped crystals and by slow cooling from 550 °C to 350 °C in 60 atm $O_2$ for the overdoped set. The nearly optimized set was grown by a self-flux method. The as-grown samples were almost optimally doped with $T_c$ = 90 K [9]. The TF-µSR experiments have been performed at the πM3 muon beam line at the Paul Scherrer Institut in Villigen, Switzerland. Spin-polarized positive muons are implanted into the bulk of the crystal. An external magnetic field $H_{ext}$ is applied perpendicular to the initial polarization of the muon spins (along the crystallographic c-axis). The muons come to rest at interstitial locations, r, randomly distributed on the length scale of the magnetic penetration depth $\lambda$. Their spins precess in the local magnetic field B(r) with the Larmor frequency $\omega_\mu = \gamma_\mu B(r)$, where $\gamma_\mu = 851.4$ MHz / T is the gyromagnetic ratio of the muon. The time evolution of the muon spin polarization P(t) is measured by monitoring the decay positrons which are preferentially emitted along the muon spin direction at the instant of decay. The probability distribution of the local magnetic field n(B), the so called '$\mu$SR-lineshape', is extracted from P(t) via Fast Fourier Transform (FFT) techniques. The real part of the FFT contains the detailed information on the vortex structure [10]. For an ordered FLL, n(B) is strongly asymmetric [10] and has a pronounced tail towards the high field side due to muons that stop near the vortex cores, a cusp which corresponds to the field at the saddle point between adjacent vortices and a cutoff on the low field side corresponding to the field minimum at the point which is most remote from the vortex cores. The asymmetry of n(B) is best characterized by the so called 'skewness' [11] $\alpha \equiv <\Delta B^3>^{1/3} / <\Delta B^2>^{1/2}$, where $<\Delta B^n>$ is the $n^{th}$ central moment of n(B). For an ordered FLL the $\alpha$-values are typically close to unity [11, 12]. The parameter $\alpha$ is very sensitive to structural changes of the vortex-state which occur, for example, as a function of increasing temperature [11,13] or magnetic field [11, 12]. The most pronounced changes have been observed in the vicinity of the irreversibility line (IL) where $\alpha$ exhibits a sudden and discontinuous change from positive to negative values [11, 13]. Previously, this marked transition has been attributed to a one-stage melting of the vortex lattice [11, 13] since no indication has been obtained for a second transition of the vortex-state in the irreversible regime below the IL.

In the following we present new TF-$\mu$SR data which show that the equilibrium vortex-state does indeed undergo a second pronounced change as a function of temperature in the irreversible regime well below the IL. Our data establish that such a second transition occurs for all three sets of under- to overdoped Bi-2212 crystals regardless of the details of their vortex mobility. Consequently, this second transition cannot be explained in terms of a crossover of the vortex dynamics or a glass transition. The unique changes of n(B) instead are consistent with a two-stage melting scenario [5] of the vortex lattice.

Fig. 1 displays the temperature dependence of the skewness α, the shift of the cusp-field with respect to the external field (cusp-shift, $B_{sh}$) and the second moment ($<\Delta B^2>$) of n(B), which have been obtained by field cooling in a magnetic field of 100 mT, 27.5 mT and 5 mT for the overdoped, the almost optimized and the underdoped crystals. The external field is always well below the dimensional crossover field H* ~ 150 mT, 60 mT and 7.5 mT, respectively [11, 12]. It was previously shown that the low temperature vortex structure undergoes a pronounced change as a function of the applied magnetic field from a regular flux-line lattice for H < H* to a strongly disordered array of decoupled pancake vortices for H > H* [11, 12]. In agreement with these previous reports, [11, 12] our low temperature 'µSR-lineshapes' at H < H* are characteristic of a FLL, i.e. we observe α ~ 1 and a fairly large value of $B_{sh}$. Also, our data confirm that n(B) exhibits the most pronounced changes in the vicinity of the IL. At the IL, which was independently determined from DC-magnetization measurements [11], the α-value drops and exhibits a sign change at $T_{IL}$ ≈ 46.5 K (0.73 $T_c$) for the overdoped, $T_{IL}$ ≈ 67.5 K (0.8 $T_c$) for the almost optimized and $T_{IL}$ ≈ 61.5 K (0.75 $T_c$) for the underdoped crystals (Fig. 1). Another characteristic feature is that $B_{sh}$ and $<\Delta B^2>$ become rather small above $T_{IL}$. In addition to these well established features, our new data show that the T-dependence of n(B) exhibits yet another noticeable anomaly which occurs in the irreversible regime at a temperature $T_m < T_{IL}$. This second transition is evident in Fig. 1 in conspicuous changes in α, $B_{sh}$ and $<\Delta B^2>$ which occur in all three sets of crystals. Furthermore, our TF-µSR measurements establish that the vortex mobility in the relevant temperature range around $T_m$ is fundamentally different for the underdoped and overdoped sets. The mobility of the vortex structure can be studied with the TF-µSR technique simply by reducing the applied field after the field-cooling process. The 'µSR lineshape' will remain unchanged if the vortices are rigidly pinned, whereas n(B) will follow the changes of the applied field if the vortices are mobile. For the underdoped set we obtain no evidence that significant vortex motion sets in below $T_m$ (not even on the time scale of hours), neither from the pinning experiments nor from the T-dependence of α and $<\Delta B^2>$. Here the sudden onset of vortex motion coincides with the transition at $T_m$. In clear contrast, for the overdoped set we observe the onset of significant vortex motion already well below $T_m$ at $T_{dp} \sim 17$ K ($0.27\, T_c$) $\ll T_m \sim 37$ K ($0.58\, T_c$). Here the signature of the vortex motion is also evident from the decrease of $<\Delta B^2>$ and the slight reduction of $\alpha$ in the vicinity of $T_{dp}$. This profound difference in the vortex dynamics of the underdoped and overdoped crystals makes us confident that the observed changes of the 'µSR lineshape' represent a unique property of the vortex lattice likely related to a two-stage melting scenario. The transition of the vortex lattice at $T_m$ thus may be associated with the intra-planar melting of the FLL into a liquid phase of flux-lines which exhibits irreversible magnetic behavior. We note that it is not possible to unambiguously assign the 'µSR lineshape' in the intermediate state for $T_m < T < T_{IL}$ to such a flux-line liquid state. Nevertheless, the existence of an intermediate phase in the (H, T) - plane at least contradicts a mere single-stage melting-scenario of the vortex structure. Furthermore, we would like to stress the fact that the (H, T) - phase diagram which emerges from our TF-µSR data (Fig. 2a - c) and complementary DC-magnetization data (Fig. 2a - c) agrees surprisingly well with the (H, T) - phase diagram which has been predicted for the two-stage melting scenario [5].

In the following we show the T-dependence of the 'µSR lineshape' in the high field regime at $H > H^*$ which provides additional support for a two-stage melting scenario. It has previously been established that the low-T vortex-state is a 'quasi-2D' solid of decoupled pancake vortices if the magnetic field exceeds a characteristic critical field $H^*$ [11, 12]. The decoupling of the pancake vortices seems to be driven by the gain in pinning energy when they become free to adjust themselves to point-like pinning sites within each individual $CuO_2$ (bi)layer. The low-T n(B) then becomes almost symmetric with strongly reduced values of $\alpha$, $B_{sh}$ and $<\Delta B^2>$ [11, 12]. Fig. 3 shows the T-dependence of $\alpha$ and $B_{sh}$ for the underdoped set at $H_{ext} = 100$ mT $> H^* = 7.5$ mT. Similarly to the low field data, the most pronounced changes of n(B) occur in the vicinity of the IL, where $\alpha$ drops and changes sign (Fig. 3) and $B_{sh}$ starts to decrease (Fig. 3). Also, a second anomaly is again apparent in the T-dependence of n(B) which occurs in the irreversible regime well below the IL. Contrary to the decrease of $\alpha$, $B_{sh}$ and $<\Delta B^2>$ at $T_m$ which occurred for $H < H^*$, the values of $\alpha$ and $B_{sh}$ now exhibit a small but significant increase. Both effects are indicative of a restoration of the inter-planar coupling as the intra-planar order of the 'quasi-2D' pancake solid diminishes above $T_m$. Particularly, the increase of $\alpha$ signals the formation of flux-line segments whose length scale (along the c-axis direction) exceeds the in-plane magnetic penetration depth $\lambda_{ab} \sim 2000$ Å [12]. Note that the increase of $\alpha$ is the more significant since in the intermediate state for $T_m < T < T_{IL}$ the thermal fluctuations and thus the vortex mobility are enhanced and tend to reduce $\alpha$. At low T the adjustment to the pinning sites was achieved by suppressing the inter-planar coupling of the individual flux-lines in favor of the persistence of 'quasi-2D' intra-planar order of the pancake vortices. At the in-plane melting transition, however, the intra-planar order is lost and there is no further need for the decoupling of the individual flux-lines in order to obtain a favorable adjustment to the pinning sites. The flux-lines are thus at least partially restored. We note that a similar trend has been observed at a higher field of $H_{ext} = 600$ mT for the sets of underdoped and overdoped Bi-2212 single crystals.

A rather interesting question is whether our observation of a two-stage melting transition truly conflicts with the numerous experimental reports of a single-stage melting transition of the vortex lattice in less anisotropic systems like Y-123. We argue that the key to the answer lies in the c-axis coupling strength, which varies widely amongst the different cuprate high-$T_c$ compounds. In contrast, the condensation energy [15] or the super-fluid density [16], which determines the in-plane vortex-vortex interaction, does not vary much. As is shown in Fig. 2d, we can simply rescale with the critical field of the dimensional crossover $H^*$ and obtain a unique (H, T) - phase diagram of the transitions at $T_m$ and $T_{IL}$ as deduced from DC-magnetization and TF-µSR measurements for a wider series of Bi-2212 single crystals. Note that recently a similar scaling property of the IL has been reported for a number of different high-$T_c$ compounds [17]. This scaling behavior implies that a two-stage melting process is experimentally observable only if H is of the order of $H^*$. In fact, it can be seen from Fig. 2d that $T_m$ and $T_{IL}$ seem to merge in the low field regime for $H \ll H^*$. It is therefore not too surprising that a single-stage transition has been observed for Y-123 which is the least anisotropic compound with $H^*$ in excess of 10 T.

In summary, we have performed muon spin rotation measurements on three sets of overdoped ($T_c = 64$ K), almost optimized ($T_c = 90$ K) and underdoped ($T_c = 77$ K) $Bi_2Sr_2CaCu_2O_{8+\delta}$ single crystals. For all three sets of crystals we observe two marked changes in the 'µSR-lineshape' as a function of temperature which can be associated with a two-stage melting transition of the vortex matter under equilibrium conditions. The well-known transition at the irreversibility line can be related to the inter-planar decoupling of the pancake vortices resulting in a pancake gas phase with reversible magnetic behavior. The second transition of the vortex-state occurs at a lower temperature $T_m$ well below the IL. We interpret this second transition as the intra-planar melting of the vortex structure into a liquid phase of flux-lines.

We would like to thank D. Herlach and A. Amato for technical support during the µSR experiments at PSI. C.T. Lin and K. Kishio are acknowledged for providing us with the crystals. The financial support of the German BMBF is gratefully acknowledged.

4-9-

Figure captions

**Fig. 1:** Data from field-cooled TF-µSR experiments on overdoped, $T_c = 64$ K (left panel), almost optimized, $T_c = 90$ K (middle panel) and underdoped, $T_c = 77$ K (right panel) Bi-2212 single crystals in an applied field of $\mu_o H_{ext} = 100$ mT, 27.5 mT and 5 mT, respecitively. Shown are the temperature dependence of the 'skewness' $\alpha$, the cusp-shift $B_{sh}$ and the second moment $<\Delta B^2>$ of n(B). $T_m$ and $T_{IL}$ denote the temperatures of the intra-planar melting and the inter-planar decoupling respectively.

**Fig. 2:** (a) - (c)(H, T) phase diagram constructed from µSR and DC-magnetization [12] data. Shown are the irreversibility field $H_{irr}$, the field for the second peak in M(H), $H_{2p}$, and for the peak in the corresponding derivative dM/dH, $H'_{2p}$ (index d for decreasing and i for increasing field). The lines are guides to the eye. (d) Rescaled DC-magnetization data (scaling factor H*) for different doping states of Bi-2212 single crystals.

**Fig. 3:** Data from field-cooled TF-µSR experiments on the underdoped Bi-2212 single crystal ($T_c = 77$ K) in an applied field of $\mu_o H_{ext} = 100$ mT. Panels are as described in Fig. 1.

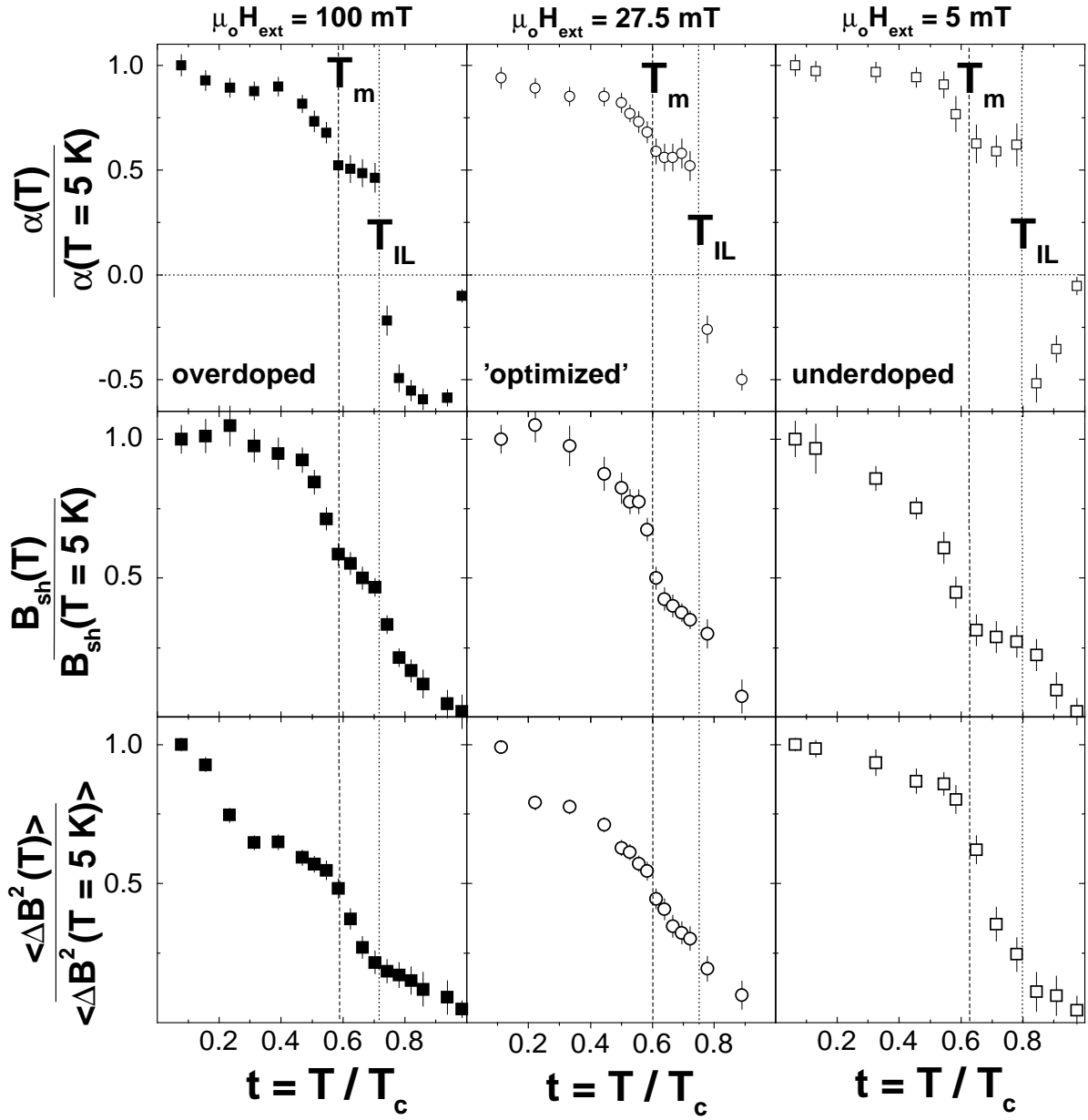

**Figure 1**
(Blasius *et al.*)

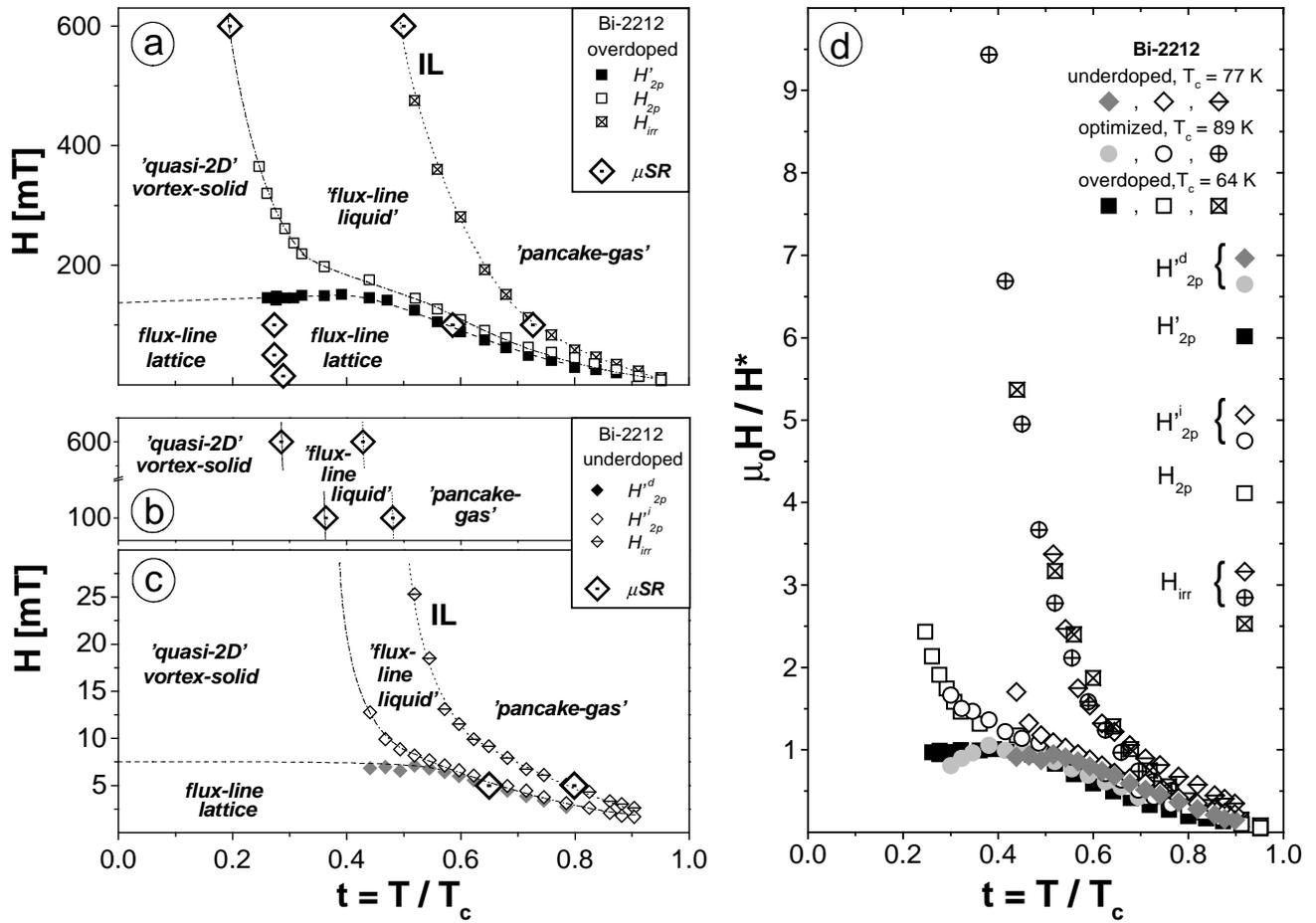

**Figure 2**
(Blasius *et al.*)

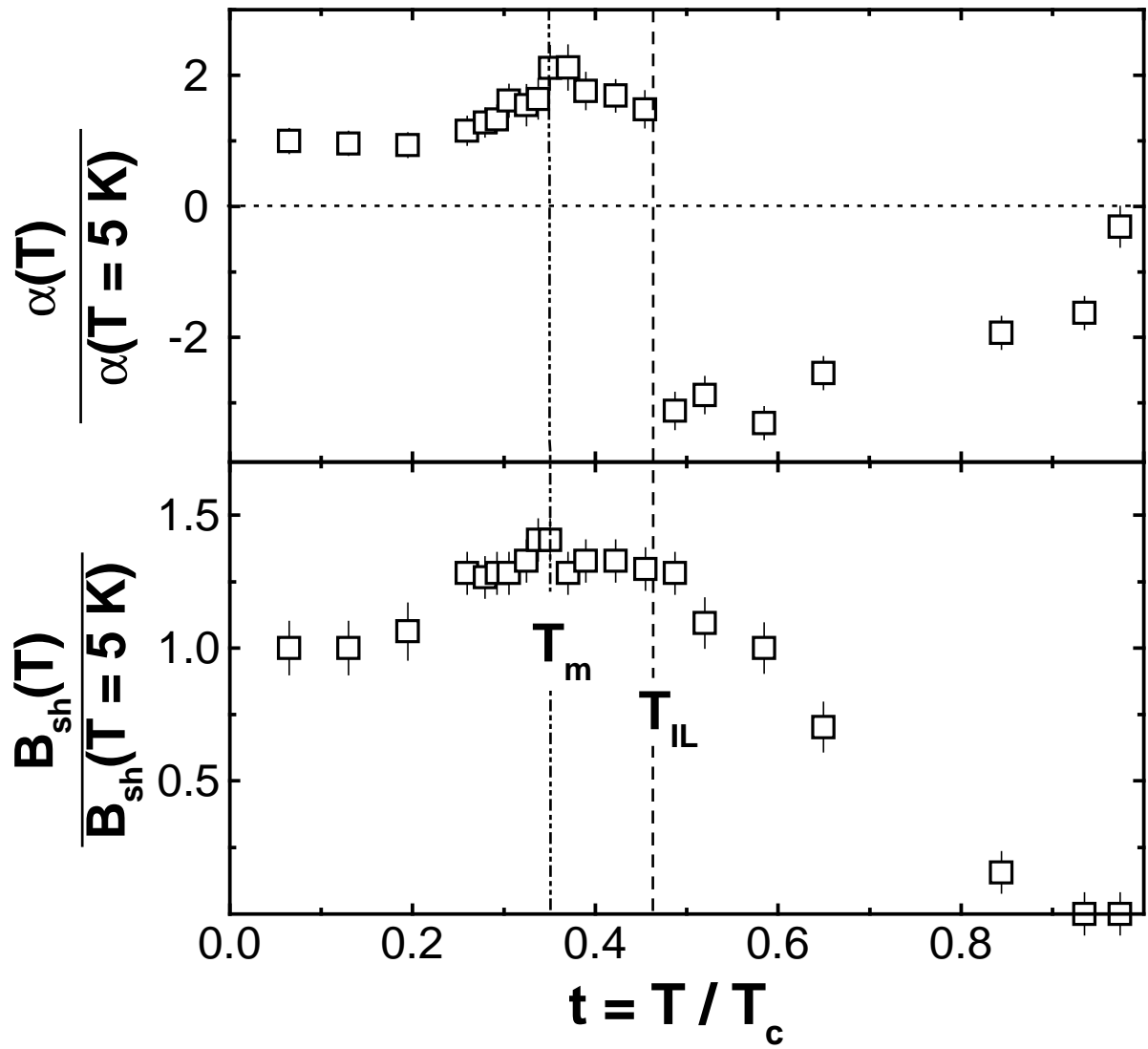

**Figure 3**
(Blasius *et al.*)